# S17 near Zero Energy in a Direct Radiative Capture Analysis

Kyung Hoon   Kim

Department of Physics Education, Sunchon National University,
Sunchon 540-742, Republic of Korea

S17 near zero energy   was calculated without using the effective expansion of the S factor or the asymptotic wave functions. Variations of the nuclear potential parameters scarcely affect the d-wave capture cross section below 0.1 MeV, but the s-wave capture cross section near zero energy is affected strongly by the shape of the nuclear potential in our calculations. This result is contrary to the existing assumption that the value of the S factor near zero energy depends on the asymptotic wave function (or asymptotic normalization coefficient). We showed   that although the s-wave contribution is dominant near zero energy, the d-wave contribution is not negligible.

The observed deficit of 8B neutrinos when compared to solar model predictions (the so-called solar neutrino problem), has focused much attention on the 7Be(p,gamma)8B reaction rate[1,2]. To date, several direct measurements of this cross section   have been performed   by using   a 7Be target[3-9].  The Coulomb dissociation of 8B [10-12] and the transfer reaction in order to determine   the asymptotic normalization constant   of the 8B   bound state wave function [1,13,14] were also used to determine the 7Be(p,gamma)8B  reaction cross section at solar energies. However it is now very difficult to get the astrophysical S factors experimentally at solar temperature.  The lowest energy  measured   to date is about 100 keV.   Therefore an extrapolation   based on some nuclear model   must be performed   to obtain the astrophysical S factors   at solar energies.  Various theoretical models have been studied   recently[15-29].   Generally, the theoretical extrapolation at very low energies can not avoid numerical difficulty because the initial state of the system   becomes so small   that its numerical determination becomes very hard   and the computation of the Coulomb functions,  which are necessary  to fix the  asymptotic normalization,  is not easy  to  deal with[26]. This numerical difficulty caused   most   of   the previous theoretical   researchers to use   the effective expansion of the S factor   or an approximation method in calculating the radial wave functions for the astrophysical S factors   near zero energy. The utilization of the transfer reaction   in order to determine the asymptotic normalization constant is one of the methods for avoiding this difficulty.

The main purposes of this work are to analyze the 7Be(p,gamma)8B reaction at low energies on the basis of a direct radiative capture mechanism and to calculate the astrophysical S factors of the 7Be(p,gamma)8B reaction near zero energy without using the effective expansion of the astrophysical S factor or the asymptotic wave function. The extranuclear capture process is so dominant at low energies that the direct radiative capture mechanism reproduces the reaction cross section fairly well [30]. From the calculations, we can check the validity of various extrapolation models. The E1, E2, and M1 transitions are considered in this theoretical research. The numerical difficulties could be overcome by using a 64-bit computer system.

The differential radiative capture cross section of the 7Be(p,gamma)8B reaction can be calculated following the formula by Kim et al.[30,32] Throughout these calculation, E1, E2, and M1 captures from $l$ = 0- 3 partial waves are considered. Various single particle 7Be + p models have been suggested and used to describe the 8B ground state and resonance. The results from microscopic calculations indicate that the overlap of the ground state of 8B with that of 7Be is very close to unity[15,27]. The quadrupole moment of 7Li is rather large, so the similarities of the properties of mirror nuclei suggests that 7Be is considerably deformed[27]. This fact means that the ground state of 8B(2+) would have a contribution from the p1/2 proton coupled to the 7Be(3/2-) channel (p1/2 * 3/2-), as well as (p3/2 * 3/2-. However, unfortunately, we have no data for the spectroscopic factor of p1/2 state in , so we assumed that the bound state of is in the single configuration 8B(2+) on the 7Be core with central and spin-orbit interactions. To reproduce the bound state wave functions, we started with the conventional bound state problem to obtain the experimental separation energy of the ground state of 8B. The parameters of the Woods-Saxon potential obtained for the ground state are Vo = 32.14 MeV, Ro= 2.95 fm, ao= 0.52 fm, and Vs.o.= 1.7 MeV, which are similar to Tombrello's [31], except for the spin-orbit interaction to be considered here.

To reproduce the radial distorted wave of incident channel, we imposed the restriction that potential parameters for the odd partial waves should reproduce the peak position of the 1+ resonance at Ec.m.= 633 keV and that the potential parameters for the even partial waves should reproduce the experimental data of astrophysical S factors at low energies. However, there are uncertainties in the experimental data at low energies, as shown in Fig. 1. To date, several direct measurements of 7Be(p,gamma)8B reaction cross section have been performed with quoted uncertainties of less than 10 %, but two of these results are larger than others, as shown in Fig.1 [1]. Therefore, we prepared two sets of potential parameters for the incident channel. The potential parameters to reproduce the experimental data of Ref. [7-9](Pot A) are Veven = 20.84 MeV, Vodd= 46.03 MeV, Ro = 2.30fm, ao= 0.65 fm, and Vs.o.= 1.7 MeV. The potential parameters to reproduce the experimental data of Ref.[4,5](Pot B) are Veven = 29.84 MeV, Vodd= 29.84 MeV, Ro = 2.95fm, ao = 0.52 fm, and Vs.o.= 1.7 MeV, which are similar to Kim's[30]. Since we were interested in very low energies, we had to be careful in generating accurate Coulomb functions and radial wave functions. For the present calculation, we adopted a 64-bit computer and program that could calculate the wave functions with high accuracy. Fig. 1 shows the calculated astrophysical S factors along with the measured ones. Our calculations using Pot. A and Pot. B reproduce well the experimental data of Refs. 7-9 and Refs. 4,5 below 0.5 MeV without any normalization factor, respectively. The calculated S values at 0.05 keV were 22.13 eV·b(Pot. A) and 24.23 eV·b(Pot. B), using the same bound state wave function for both cases. This is a striking result, contrary to the existing assumption that the S value near zero energy depends on the asymptotic wave function (or asymptotic normalization coefficient). We could confirm this potential dependency of S factor from the fact that the trial potential set (Pot. C) for the incident channel, Veven= 59.84 MeV, Vodd = 29.84 MeV, Ro = 2.95fm, ao = 0.52 fm, and Vs.o.= 1.7 MeV, gave an S value at 0.05 keV of 9.39 eV·b when the same bound state function was used. The asymptotic condition of $k*r \gg 1$ in continuum state means that if k goes to zero, then r goes to infinity. Therefore the overlap integration in non-asymptotic region is important at very low energy. This is shown in Fig. 2.

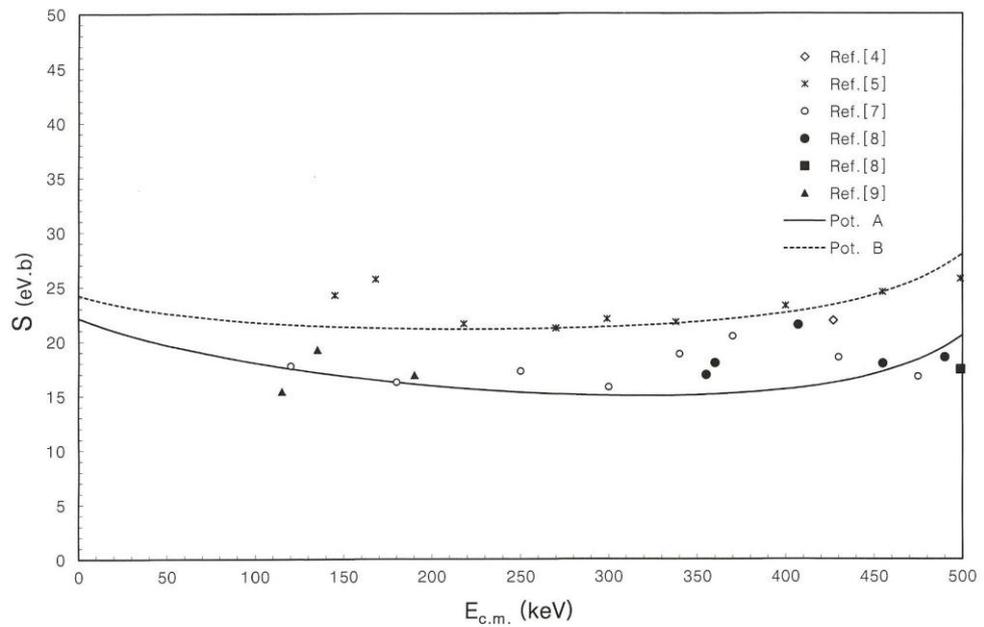

Fig. 1. Calculated astrophysical S factor along with experimental data.

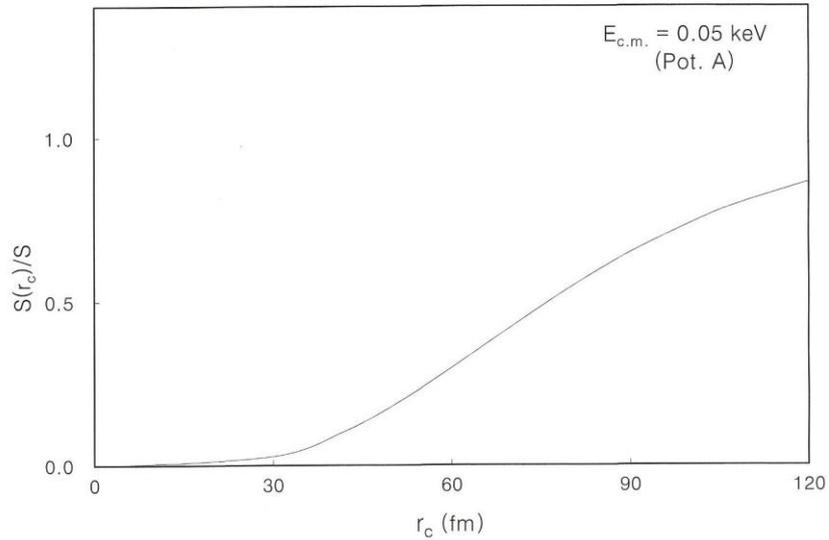

Fig.2. $S(r_c)$ is the S value when the overlap integral is integrated from zero to $r_c$.

We checked that E2 and M1 captures lead to negligible contributions below 100 keV in this reaction. The E1 transitions to the 2+ ground state are dominated by the s wave capture at very low energies, but the d wave contribution is not negligible. This is clearly shown in Fig .3 where we display the ratio of d wave contribution to the cross section leading to the ground state. We checked that in our calculations, the variations in the potential parameters scarcely affected the d wave capture cross section below 0.1 MeV. However the s wave capture cross section near zero energy was affected strongly by the shape of nuclear potential in this reaction. The calculated s wave capture cross sections at 0.05 keV are $18.3 * 10(-223)$ μb (Pot.A), $20.2 * 10(-223)$ μb (Pot.B), and $6.98 * 10(-223)$ μb (Pot.C) while the d wave capture cross section has the same value of $1.34 * 10(-223)$ μb for all cases. We attained calculated S values at 20 keV of 21.0 eV·b (Pot.A) and 23.4 eV·b (Pot.B) by using the same bound state wave function for both cases in this single particle model study.

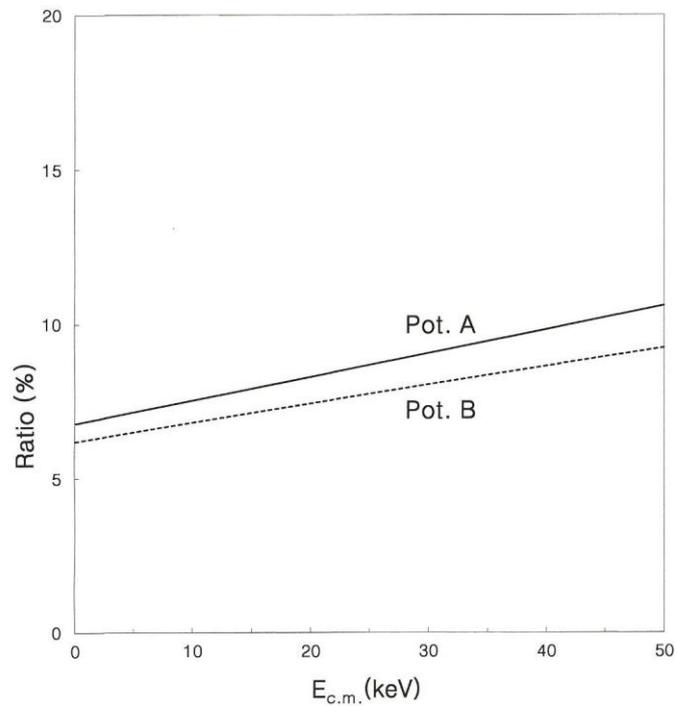

Fig. 3. The ratio of d wave contribution to the cross section leading to the ground state.